\documentclass[useAMS,usenatbib,usegraphicx]{mn2e}

\usepackage{bm}

\title[The feeble giant. Discovery of a large and diffuse Milky Way dwarf galaxy in Crater] {The feeble giant. Discovery of a large and diffuse Milky Way dwarf galaxy in the constellation of Crater\thanks{Based on data products from observations made
    with ESO Telescopes at the La Silla Paranal Observatory under
    public survey programme ID programme 177.A-3011(A,B,C)}}

\author[G.~Torrealba et al.]
{G.~Torrealba$^1$,
  S.E.~Koposov$^1$,
  V.~Belokurov$^1$ 
  \& M. Irwin$^1$
\\ $^1$Institute of Astronomy, Madingley Rd, Cambridge, CB3 0HA
}
\begin{document}

\pagerange{\pageref{firstpage}--\pageref{lastpage}} \pubyear{2015}

\maketitle

\label{firstpage}

\begin{abstract}

We announce the discovery of the Crater~2 dwarf galaxy, identified in imaging data of the VST ATLAS survey. Given its half-light radius of $\sim$1100 pc, Crater~2 is the fourth largest satellite of the Milky Way, surpassed only by the LMC, SMC and the Sgr dwarf. With a total luminosity of $M_V\approx-8$, this galaxy is also one of the lowest surface brightness dwarfs. Falling under the nominal detection boundary of 30 mag arcsec$^{-2}$, it compares in nebulosity to the recently discovered Tuc~2 and Tuc~IV and UMa~II. Crater~2 is located $\sim$120\,kpc from the Sun and appears to be aligned in 3-D with the enigmatic globular cluster Crater, the pair of ultra-faint dwarfs Leo~IV and Leo~V and the classical dwarf Leo~II. We argue that such arrangement is probably not accidental and, in fact, can be viewed as the evidence for the accretion of the Crater-Leo group.
\end{abstract}


\section{Introduction}\label{sec:INTRO}

The size and the luminosity of a dwarf galaxy satellite today are stipulated by the star-formation efficiency at birth and the amount of host harassment it endures during the rest of its life. Therefore, local dwarfs can only be used to scrutinize the high-redshift structure formation in the low-mass regime if the effects of the host influence can be singled out. In principle, such inference should be possible if the satellite's orbital history is known. In practice, unfortunately, this is easier said than done, both due to the challenging nature of the proper motion measurement, as well as the unexplored effects of dynamical friction and orbital fanning. It might, however, be possible to decode the satellite's orbital history if it was accreted as part of a group, thus linking its observed properties to the environment at the origin.

Simulations suggest that half of all satellites at $z=0$ were in groups before falling into Milky Way-like hosts \citep{Wetzel2015}. Therefore, in any realistic Galactic assembly history, it is unlikely that the current satellite distribution will look isotropic. Indeed, the apparent spatial anisotropy of the Galactic satellite population has been pointed out on a number of occasions \citep[see e.g.][]{Lyndenbell1976, Kroupa2005,Pawlowski2012}, as well as its alignment with the Magellanic Clouds orbit \citep[see e.g.][]{LyndenBell1995}. As \citet{Deason2015} point out, the spatial extent of an accreted group depends on its mass and the time of its in-fall into the Milky Way, with the least massive (and hence the coldest) and the latest events standing out the most. In this picture, the discovery last year of a large number of faint satellites near the Magellanic Clouds \citep[see][]{Koposov2015,Bechtol2015,DrlicaWagner2015} is naturally explained with a relatively recent accretion of a massive Magellanic system.

\begin{figure*}
    \includegraphics[width=\textwidth]{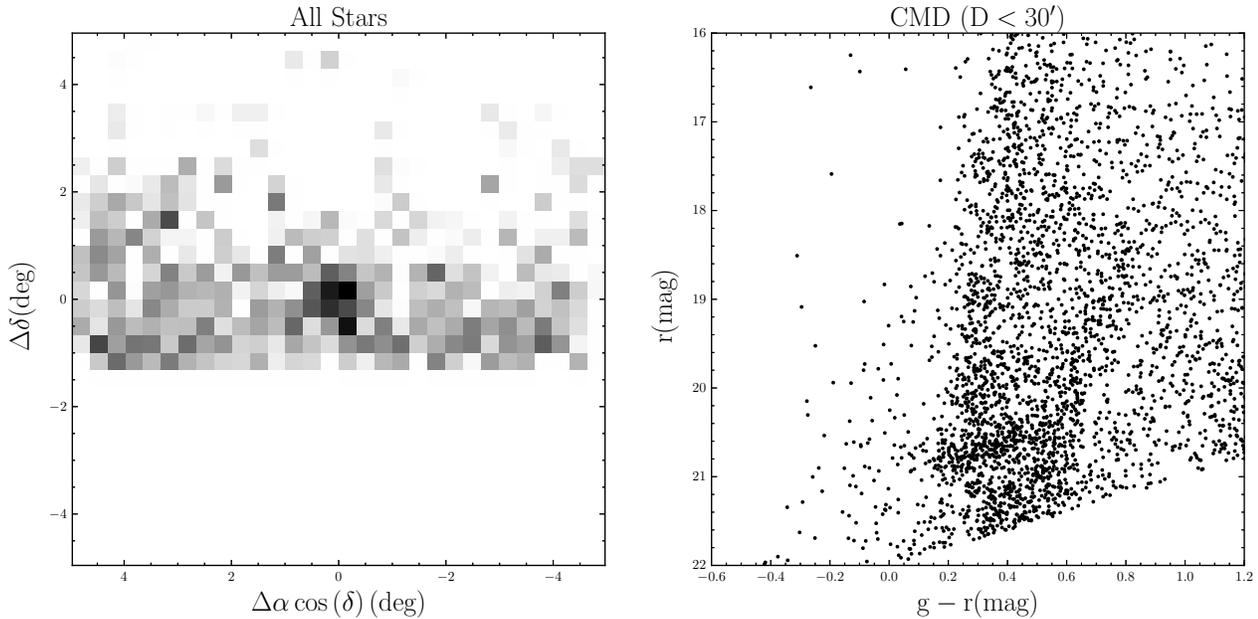}
    \caption{
      Crater~2 as seen in the ATLAS data. The left panel shows the density map on the sky of all the stars in a 10 $\times$ 10 degree area around the center of the detection. The overdensity, which is easily visible at the center, is at the border of the ATLAS footprint embedded in a non-uniform background. In the right panel, we show the CMD of all the stars within 31\arcmin (the half-light radius) of the center of the detection. Even with the strong background gradient, a prominent red giant branch and red horizontal branch stand out.}
    \label{fig:DiscoveryRaw}
\end{figure*}

However, it remains unclear the hypothesis of a recent Magellanic accretion can explain both the satellites in the immediate vicinity of the Magellanic Clouds, such as those picked out by the Dark Energy Survey (DES), {\it as well as} satellites like Draco and Ursa Minor which are hundreds of degrees, and kpc, away \citep[see e.g.][]{Nichols2011,Sales2011}. One possibility is that the group was a lax association initially and fell in earlier, losing some of its outer parts to the Milky Way tides before the LMC-SMC and its entourage were eventually captured \citep[see also][Jethwa et al. (in preparation)]{Yozin2015}. It is curious, nonetheless, that given the stipulated ubiquity of group environments, apart from the obvious LMC-SMC connection and their likely bond with many of the DES dwarfs, very few other satellite associations are known in the Milky Way. For example, it is quite possible that, being the third most massive satellite galaxy, the Sgr dSph \citep[][]{Ibata1994,MNO2010} could have had smaller dwarf companions, but only one possible candidate was found in its propinquity \citep[][]{Laevens2015}. Another plausible satellite pair are two of the Leos, IV and V \citep[][]{Belokurov2007,Belokurov2008}. What makes their connection conceivable is the combination of their proximity on the sky, i.e.$<3^{\circ}$ and small differences in their line-of-sight distances and velocities, $<$20\,kpc and $<$40\,km s$^{-1}$ respectively. Such closeness is striking given that the outer halo of the Milky Way is mostly empty, with only $\sim$20\% of all currently known dwarfs inside a 400\,kpc radius located beyond 150\,kpc from the Galactic centre.

For most accretion event architectures, the satellites that fell into the Galaxy together would share the orbital motion and hence should be found near a great circle \citep[see][]{LyndenBell1995}. Interestingly, a peculiar globular cluster, Crater \citep[][]{Belokurov2014,Laevens2014} lies very close to the great circle passing through Leo~IV and Leo~V \citep[][]{Belokurov2014}. Crater holds the record for the most distant Milky Way cluster, at 145\,kpc. It is also surprisingly young with an age of 7.5 Gyr, and possesses a Horizontal Branch that is atypically red and short for its metallicity of [Fe/H]=-1.65 \citep[][]{Weisz2015}. Finally, Crater is one of the largest globular clusters in the Galaxy. When compared to the rest of the Galactic GCs, Crater peels off as an obvious outlier. \citet{Weisz2015} conjecture that Crater could possibly originate in the SMC, even though it would not have been the most representative cluster in its former host. The fact that the Crater-Leo~IV/V great circle with the pole at $(\alpha,\delta)=(83^{\circ}.1, −5^{\circ}.3)$ \citep[][]{Belokurov2014} passes only few degrees away from the Magellanic Stream $(\alpha,\delta)=(84^{\circ}.3, 17^{\circ}.8)$\citep[][]{Nidever2008} lends further support to a possible Magellanic origin hypothesis.

In this paper we present the discovery of a new dwarf galaxy in the constellation of Crater detected using the most recent VST ATLAS data. Crater~2 appears to be one of the lowest surface brightness Milky Way dwarfs, as well as one of the largest. Moreover, there exists a great circle passing through Crater~2, Crater, Leo~IV, Leo~V and Leo~II with a standard deviation of only 0.5 degrees. This paper is organized as follows. Section \ref{sec:Discovery} introduces the VST ATLAS survey and gives the details of the discovery of the satellite. Section \ref{sec:prop} describes the in-depth stellar populations and structural properties modelling of the system. Finally, Section \ref{sec:CONCL} puts the discovery into context and assesses the plausibility of the Crater-Leo group's existence.

\begin{figure*}
    \includegraphics[width=\textwidth]{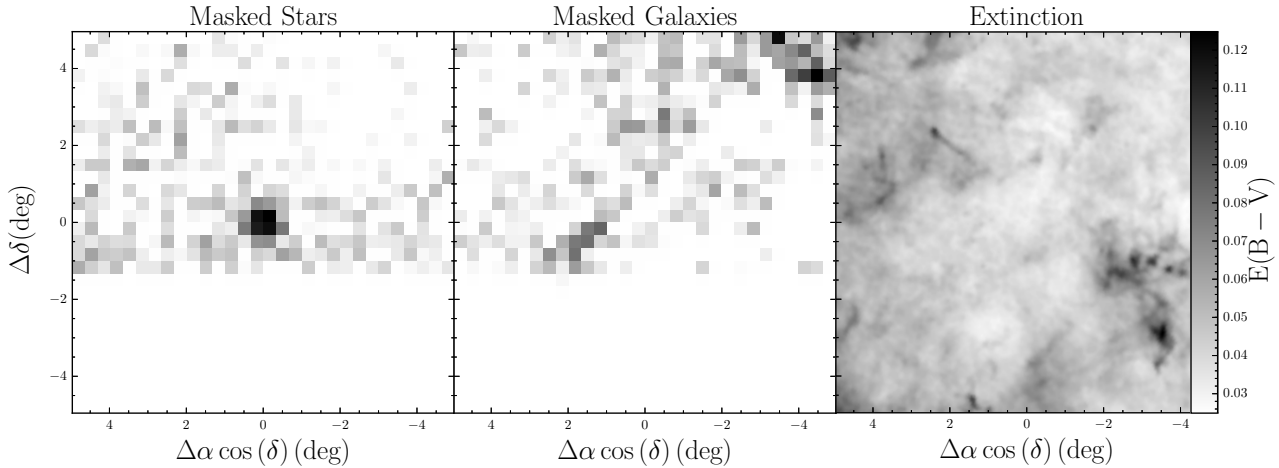}
    \caption{
      Density maps of the region around Crater~2. The left panel shows the distribution of stars filtered by an isochrone mask with [Fe/H]=-1.7, age of 10\,Gyr and m$-$M=20.35, which shows a significant enhancement of the overdensity when compared to Figure~\ref{fig:DiscoveryRaw}. The middle panel shows the density map of objects morphologically classified as galaxies in the same area, and the right panel shows the extinction map from \citet{SFD}. It is reassuring to see that there is neither an overdensity in galaxies nor strong extinction variation associated with Crater~2.}
    \label{fig:DiscoverySpatial}
\end{figure*}

\section{Discovery of Crater~2}\label{sec:Discovery}

\subsection{VST ATLAS}

ATLAS \citep{Shanks2015} is one of the three public ESO surveys currently being carried out by the a 2.6m VLT Survey Telescope (VST) at the Paranal observatory. The VST is equipped with the 16k $\times$ 16k pixels CCD camera OmegaCAM, which provides 1-degree field of view with a resolution of 0.\arcsec21 per pixel. ATLAS aims to survey 4500 square degrees of the Southern celestial hemisphere in 5 photometric bands, $ugriz$, with depths comparable to the Sloan Digital Sky Survey (SDSS). The median limiting magnitudes, corresponding to the $5\sigma$ source detection limits, are approximately 21.99, 23.14, 22.67, 21.99, 20.87 for each of the $ugriz$, respectively. Image reduction and initial catalog generation are performed by the Cambridge Astronomical Survey Unit (CASU) \citep[see][for details]{Koposov2014}. The band-merging and selection of primary sources were performed as separate steps using a local SQL database. To improve the uniformity of the photometric calibration of the survey, on top of the nightly zero-points measured relative to APASS survey, we also applied an additional global calibration step \citep[a.k.a. uber-calibration;][]{Padma08}. In this work, we use the photometric catalogs provided by CASU, which include the entirety of the ATLAS data taken up to September 2015 covering $\sim$ 4500 square degrees in at least one band, and with $\sim$3500 square degrees having both $g$ and $r$ band observations. In the analysis that follows we correct ATLAS photometry for the effects of Galactic extinction using the \citet{SFD} maps and the extinction coefficients from \citet{Yuan2013}.

\subsection{Discovery}

We trawled through the ATLAS data using a version of the systematic overdensity detection algorithm \citep[see e.g.][]{Koposov2008, Koposov2015}. Briefly, the satellite detection algorithm starts by filtering stars using an isochrone mask for a given age, metallicity and distance. The local density of the thus selected stars is then measured and compared to the density level on much larger scales, i.e. the Galactic background/foreground. In practice, the overdensity estimate is performed by convolving the masked stellar number count distribution with a ``Mexican hat'' kernel: a difference between a narrow inner kernel (for the local density estimation) and a wide outer kernel (to gauge the background density). In our implementation, both kernels are two-dimensional Gaussians and the significance of the detection at each pixel is calculated by comparing the result of the convolution with the expected variance.

\begin{table}
    \caption{Properties of Crater~2}
    \centering
    \label{tab:Properties}
    \begin{tabular}{@{}lrl}
        \hline
        Property               & Value                    & Unit\\
        \hline
        $\alpha ({\rm J2000})$ & $177.310  \pm 0.03$      & deg \\ 
        $\delta ({\rm J2000})$ & $-18.413 \pm 0.03$       & deg  \\ 
        m$-$M                  & $20.35 \pm 0.02$         & mag\\ 
        D$_\odot$              & $117.5 \pm 1.1$          & kpc\\ 
        r$_{h}$                & $31.2 \pm 2.5$           & arcmin\\ 
        r$_{h}$                & $1066 \pm 84$            & pc\\ 
        M$_{\rm V}$            & $-8.2 \pm 0.1$           & mag\\ 
        1$-$b/a                & $<$0.1 (95\%)            &\\
        age                    & $10 \pm 1$               & Gyr \\
        $[$Fe/H$]$             & $-1.7 \pm 0.1$           & dex \\
        $<$$\mu$$>$(r$<$r$_h$) & $30.6 \pm 0.2$           & $\rm{mag}\,\rm{arcsec}^{-2}$ \\
        \hline
    \end{tabular}
\end{table}

\begin{figure*}
    \includegraphics[width=\textwidth]{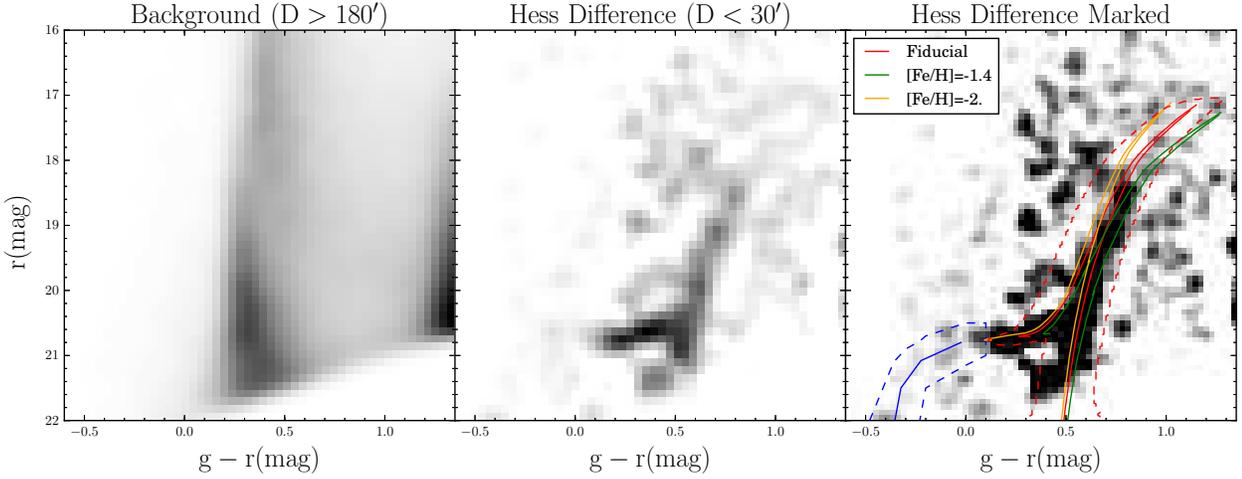}
    \caption{
      Colour-Magnitude Diagram. The left panel shows the comparison Hess CMD constructed using stars between 3 and 5 degrees away from Crater~2. The middle panel shows the Hess difference between the stars within 31\arcmin of the center and the comparison displayed in the left panel. The RGB and the HB of Crater~2 are now unmistakable. The right panel shows a contrast enhanced version of the middle panel and highlights the possible existence of a blue horizontal branch. The red line correspond to a PARSEC isochrone with [Fe/H]=-1.7, age $10$ Gyr and m$-$M=20.35. The associated mask is shown with the red dashed line. For comparison, isochrones with $[$Fe/H$]$= -1.4 and -2 are also overlaid in green and orange color respectively. The blue lines show the M92's BHB ridge-line shifted to the distance modulus of $m-M=20.35$.}
    \label{fig:DiscoveryCMD}
\end{figure*}

We applied the above method to the ATLAS data using a grid of inner kernel sizes (from 1\arcmin\ to 10\arcmin) and isochrone masks to create a list of stellar overdensity candidates across a range of stellar and structural parameters. Our isochrone masks are based on the most recent PARSEC evolutionary models \citep{Bressan2012}, which are convolved with the appropriate ATLAS photometric error. We have probed $15<m-M<23$ in distance modulus (DM), from 9.5 to 10.1 in logarithm of age, and $-2<[$Fe/H$]<-1$ in metallicity. Once the known overdensities (previously detected Galactic satellites, Local Volume galaxies etc.) are culled, in the detections list, a large object in the constellation of Crater stands out at the highest significance of $10.7\,\sigma$. As shown in Figure \ref{fig:DiscoveryRaw}, Crater~2 reveals itself rather unambiguously in both stellar spatial distribution - even before any isochrone filtering is applied - and the Color Magnitude Diagram (CMD). The left panel of the Figure shows Crater~2 as an extended overdensity at the edge of the ATLAS footprint on top of a noticeably non-uniform background. The Figure's right panel shows the CMD of all stars within $\sim30\arcmin$ of the center of the detection (see Section~\ref{sec:spatpar} for details of the measurement of Crater~2 position, distance and size), where a prominent Red Horizontal Branch (RHB) as well as an obvious Red Giant Branch (RGB) can be seen. The well-defined RHB is a decent distance indicator: Crater~2 is $\sim$120\,kpc away from the Sun. Note that curiously, Crater 2's CMD is remarkably similar to that of Sextans \citep[see Figure~3 of][]{Sextans}.

Figure \ref{fig:DiscoverySpatial} shows the density distribution of stars selected using a mask based on a PARSEC isochrone with [Fe/H]=$-1.7$, age $10$ Gyr offset to $m-M=20.35$ (see Section \ref{sec:isopar} for details on the selection of the isochrone parameters, and the right panel of Figure \ref{fig:DiscoveryCMD}). The mask's width is defined by the ATLAS photometric error above the minimum width of 0.1 mag. The CMD selection enhances the stellar overdensity associated with Crater~2 as revealed in the left panel of Figure \ref{fig:DiscoverySpatial}. According to the middle and right panels of the Figure, no obvious galaxy overdensity or strong extinction variation can be seen overlapping with the new satellite's location. Crater~2's angular diameter of $\sim$ 1 degree corresponds to $\sim$ 2\,kpc at the distance of 120\,kpc, which places it firmly in the class of dwarf galaxies.

Due to the substantial size of Crater~2, and hence the large aperture required to create the CMD, the details of the satellite's stellar population distribution are buried in the Galactic background/foreground. As a simple but efficient way of decontaminating the CMD, Figure~\ref{fig:DiscoveryCMD} (middle panel) presents the Hess difference diagram of the stars within 31\arcmin\ of the center of Crater~2, in which both the RGB and HB are now evident. For comparison, on left panel we also show the Hess diagram of the background, which is constructed with the stars that lie between 3 and 5 degrees away from Crater~2. The right panel of the Figure gives an alternative, highly contrasted version of the Hess difference which highlights hints of a faint Blue HB. Also shown is the best-fit isochrone with its associated mask in red solid line and red dashed line, respectively. For comparison, models with higher and lower metallicity are given. Additionally, the M92 BHB ridge-line shifted to the fiducial DM of $m-M=20.35$ and the BHB mask are displayed in blue.

\section{Properties of Crater~2}\label{sec:prop}
\subsection{Stellar populations}\label{sec:isopar}

To estimate the age, metallicity and distance of Crater~2 we model the distribution of the stars in the color and magnitude space with a combination of an empirical background CMD density and a single stellar population model from the PARSEC isochrone set with the appropriate photometric errors applied. The probability of observing a star with colour $g-r$ and apparent magnitude $r$ in the vicinity of Crater~2 is:

\begin{equation}\label{cmdPDF}
P(g-r,r|\phi)=f\,P_{obj}(g-r,r|\phi)+(1-f)\,P_{bg}(g-r,r)
\end{equation}

\noindent Here $\phi$ are the three isochrone parameters, i.e., age, metallicity and the distance modulus, $f$ is the fraction of stars belonging to the object, $P_{obj}$ the probability distribution of the satellite stars, and $P_{bg}$ the probability distribution of the background stars. Note that the background model, $P_{bg}$, does not depend on any model parameters and is determined empirically using the stellar distribution far from the center of the satellite. The satellite model, $P_{obj}$, is a convolution of the expected number of stars at each point along the isochrone with the typical observed photometric error of the ATLAS survey at the given magnitude.

For stars within 31\arcmin\ of the center of Crater~2, the likelihood is sampled using the affine invariant ensemble sampler {\it emcee} \citep{GoodmanWeare2010,ForemanMackey2013} with flat priors on all three isochrone parameters. Given the strong RGB and RHB of Crater~2, all three are well constrained: [Fe/H]=$-1.7\pm0.1$; age=$10 \pm 1$ Gyr and m$-$M=$20.35\pm0.02$ (also see Table \ref{tab:Properties}). Note, however, that these numbers, and in particular the formal error on the distance, should be taken with caution, given that the behavior of HB stars is still not well understood, therefore rendering the models somewhat uncertain \citep[see e.g.][]{Gratton2010}. Accordingly, an order of magnitude for the additional uncertainty in the distance can be drawn from the uncertainties of the absolute magnitudes of RHB stars, which are of the order of $\sim0.1$ mag \citep{Chen2009}, or $\sim5\%$ in the distance.

\subsection{Structural parameters}\label{sec:spatpar}

\begin{figure*}
    \includegraphics[width=\textwidth]{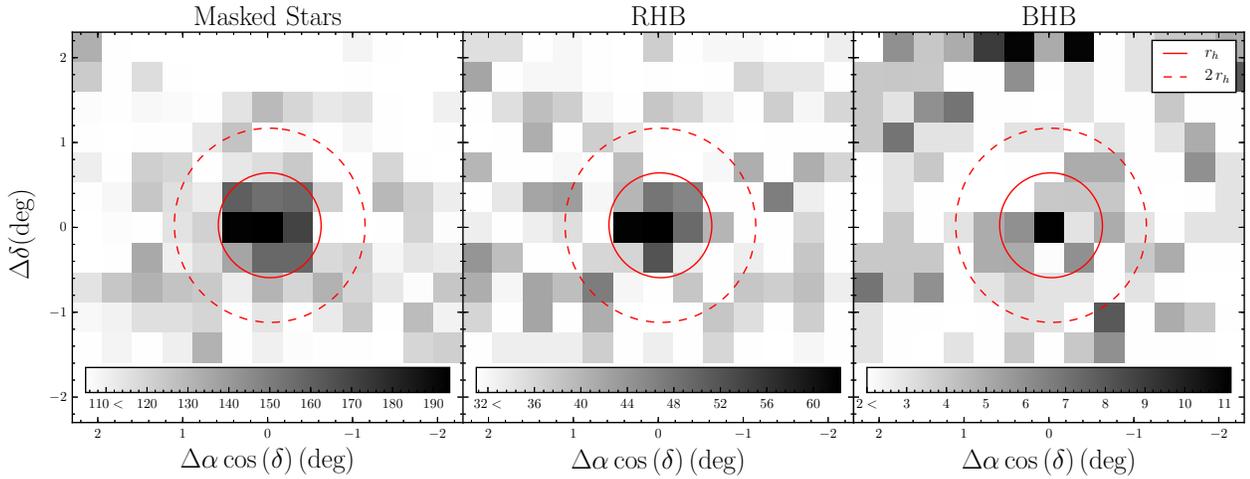}
    \caption{
      Close-up of the spatial distribution of Crater~2. The left panel shows the density map of all stars that fall within the isochrone mask, i.e. RGB + RHB. The middle panel shows RHB stars only - these are the most numerous stars above the limiting magnitude of the ATLAS survey. The rightmost panel gives the distribution of the BHB candidate stars as selected using the blue box shown in the right panel of Figure~\ref{fig:DiscoveryCMD}. The red solid (dashed) line indicates the half-light radius (twice the half-light radius), and the color scale in the bottom shows the number of stars per bin.}
    \label{fig:SpatialZoom}
\end{figure*}

The structural parameters of Crater~2 are determined by maximising the likelihood of a model describing the distribution of stars in the vicinity of the object \citep[similar to e.g.,][]{Martin2008,Koposov2015}. To reduce Galactic contamination, we only use stars with $g<22$ and $r<22$ that fall inside the isochrone mask shown in the right panel of Figure~\ref{fig:DiscoveryCMD}. The spatial density model is a combination of a linearly varying background and a 2-dimensional elliptical Plummer profile, defined as:

\begin{equation}
P_{obj}(x,y|\Theta)=\frac{1}{\pi a^2 \left(1-e\right)}\left(1+\frac{\tilde{r}^2}{a^2}\right)^ {-2},
\end{equation}

\noindent where $x$ and $y$ are the coordinates of the stars on the sky in the tangential projection at the center of the object. $\Theta$ is a shorthand for all model parameters, namely the elliptical radius $\tilde {r}$, the coordinates of the the center of Crater~2 $x_0$, $y_0$, the positional angle of the major axis $\theta$ and the ellipticity of the object $e$:

\begin{figure}
    \includegraphics[width=0.98\columnwidth]{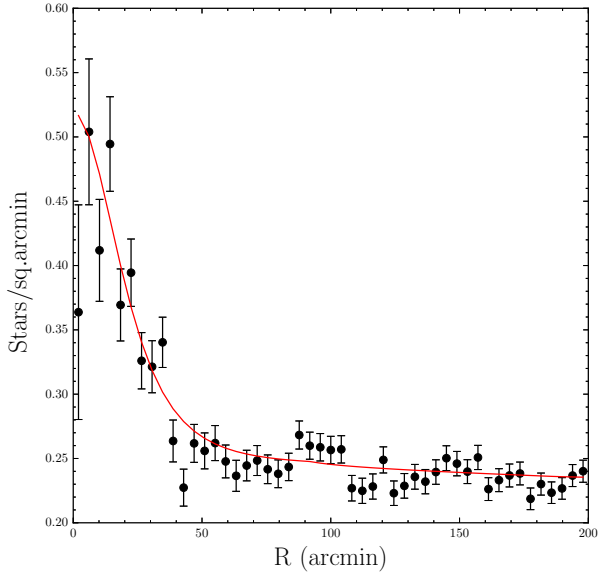}
    \caption{
      Radial number density profile of Crater~2 as probed by the stars inside the isochrone mask shown in Figure~\ref{fig:DiscoveryCMD}. The red line shows the best fit model which includes a Plummer profile with $r_h=31\arcmin$. The small bump at $R\sim100\arcmin$ is probably associated with the small overdensity of mis-classified galaxies located close to Crater~2.}
    \label{fig:RadialProfile}
\end{figure}
\begin{eqnarray}
\tilde{r}&=&\sqrt{\tilde{x}^2+\tilde{y}^2} \\
\tilde{x}&=&\cos\theta\,\left(x-x_0\right)/\left(1-e\right)+\sin\theta\left(y-y_0\right)/\left(1-e\right) \nonumber\\
\tilde{y}&=&-\sin\theta\,\left(x-x_0\right)+\cos\theta\left(y-y_0\right)\nonumber.
\end{eqnarray}

\noindent The background/foreground density is a bilinear
distribution of the form:

\begin{eqnarray}
P_{bg}(x,y|\Theta)=\frac{1}{N_{bg}}\left(p_1\,x+p_2\,y+1\right),
\end{eqnarray}
\begin{figure*}
    \includegraphics[width=\textwidth]{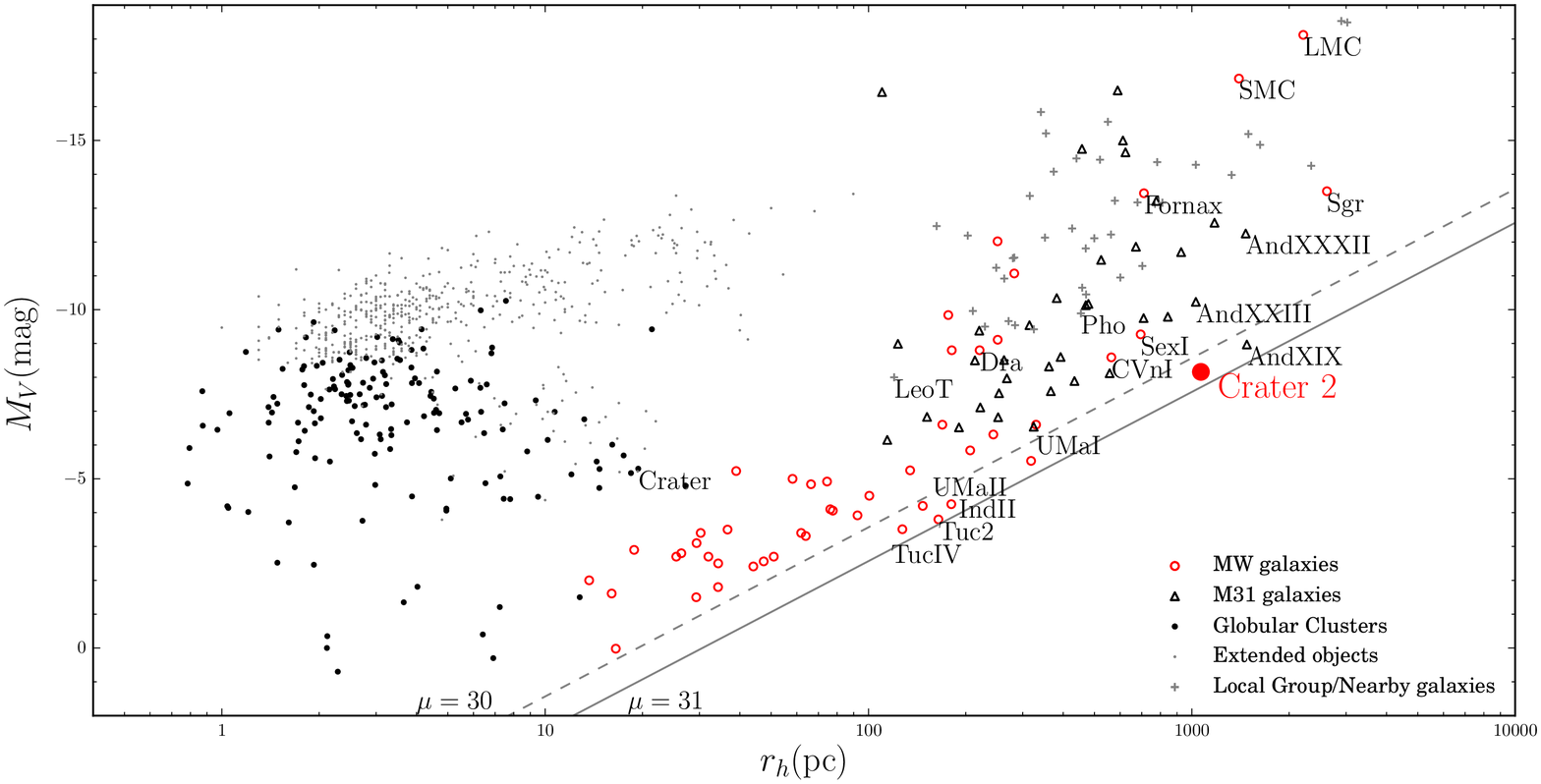}
    \caption{
      Absolute magnitude versus half-light radius. Local galaxies from \citet{McConnachie2012} (updated September 2015) are shown with different symbols. Dwarf galaxy satellites of the Milky Way are shown with red open circles, the M31 dwarfs with black unfilled triangles, and other nearby galaxies with grey crosses. Black dots are the MW globular clusters measurements from \citep{Harris2010,Belokurov2010,Munoz2012,Balbinot2013,Kim2015a,Kim2015b,Kim2015c,Laevens2015,Weisz2015}  and grey dots are extended objects smaller than 100\,pc from \citep{Brodie2011}. As the half-light radii of LMC and SMC are unavailable in \citet{McConnachie2012}, we measured these ourselves using the 2MASS point source catalogue \citep{Skrutskie2006}. The black solid (dashed) line corresponds to the constant level surface brightness within half-light radius of $\mu=31$ (30) $\rm{mag}\,\rm{arcsec}^{-2}$, which is approximately the surface brightness limit of the searches for resolved stellar systems in the SDSS \citep{Koposov2008}. The position of Crater~2 is marked with a a filled red circle. Apart from the only three MW dwarfs exceeding Crater~2 in size, i.e. the LMC, the SMC and the Sgr, we also label the following systems. UMa~I, UMa~II, Tuc~2 and Tuc~IV are the ultra-faint dwarfs with surface brightness levels similar to that of Crater 2. Leo T, Dra, Pho, Sex and CVn I all have similar (or slightly higher luminosity) but are smaller in size. Fornax is overwhelmingly more luminous compared to Crater 2, yet not as extended. Finally, there are three systems in the M31 that are comparable (or even larger!) in size to Crater~2: And XIX, And XXIII and And XXXII. The position of the peculiar and extended globular cluster Crater is also marked.}
    \label{fig:SizeLuminosity}
\end{figure*}

\noindent with $p_1$ and $p_2$ the two parameters that define the plane, and $N_{bg}$ a normalization constant so the integral of $P_{bg}$ over the modeled area is 1. Analogously to Eq~\ref{cmdPDF}, the probability of observing a star at $x,y$ is:

\begin{equation}
P(x,y|\Theta)=f\,P_{obj}(x,y|\Theta)+(1-f)\,P_{bg}(x,y|\Theta),
\end{equation}

\noindent where $f$ is the fraction of stars belonging to the object.

As in the case of the CMD modelling, flat priors are assumed for all parameters, and the posterior distribution is sampled with {\it emcee}. Here we report the best fit parameters and their uncertainties corresponding to the 15.9, 50, and 84.1 percentiles of the posterior distribution (see Table \ref{tab:Properties}). Interestingly, we find that in projection on the sky, Crater~2 appears to be completely circular with e$<$0.1 at 95\% confidence. Its half-light radius is $31'\pm2.5'$, corresponding to the physical size of 1.07$\pm$0.08\,kpc at the measured distance of 117\,kpc (m$-$M=20.35). A zoom-in on the spatial distribution of stars in the proximity of Crater~2 is shown in Figure \ref{fig:SpatialZoom}. The left panel displays the density map of stars that belong to both the RHB and the RGB, the middle panel presents the RHB candidates only, while the right panel deals with BHB density. The BHB candidate stars are selected to lie within the blue dashed box shown in the right panel of Figure~\ref{fig:DiscoveryCMD}. The red solid (dashed) circle shows the half-light radius $r_h$ ($2r_h$) of the best-fit model. An overdensity of stars is visible in both middle and right panel of Figure \ref{fig:SpatialZoom}, with the RHB signal being clearly stronger. The radial profile of Crater~2 is shown in Figure~\ref{fig:RadialProfile} together with the best fit model (red line), which clearly provides an adequate fit to the data. The small bump at $R\sim100'$ is probably caused by the contamination from mis-classified galaxies residing in the galaxy overdensity seen in Figure \ref{fig:DiscoverySpatial}.

The total number of Crater~2 stars with $g,r<22$ inside the isochrone mask can be obtained from the measurement of the fraction of Crater~2 member stars in the field $f$. Using the total number of stars $822\pm 79$, we compute the absolute magnitude of Crater~2 by assuming a Chabrier initial mass function \citep{Chabrier2003} and the best-fit PARSEC isochrone described in the previous section. The above calculation gives the intrinsic luminosity of the satellite of $M_V = -8.2\pm 0.1$.

\section{Discussion and Conclusions}\label{sec:CONCL}
\subsection{The feeble giant}

Figure \ref{fig:SizeLuminosity} compares the structural properties of Crater~2 to those of known satellite systems. Estimates of half-light radius are shown against the values of absolute magnitude for all currently known Milky Way globular clusters (black dots) and dwarf galaxies (open circles) as well as the dwarfs of the M31 (triangles) \citep[][updated Sep 2015]{McConnachie2012}; this is complemented by measurements collected for various extended companion objects around nearby galaxies \citep[grey dots,][]{Brodie2011}. To guide the eye, the lines of constant surface brightness of 30 and 31\,mag\,arcsec$^{-2}$ is drawn. This level - as deduced by e.g. \citet{Koposov2008} - approximately corresponds to the surface brightness detection limit of the current dwarf satellite searches.

\begin{figure*}
    \includegraphics[width=\textwidth]{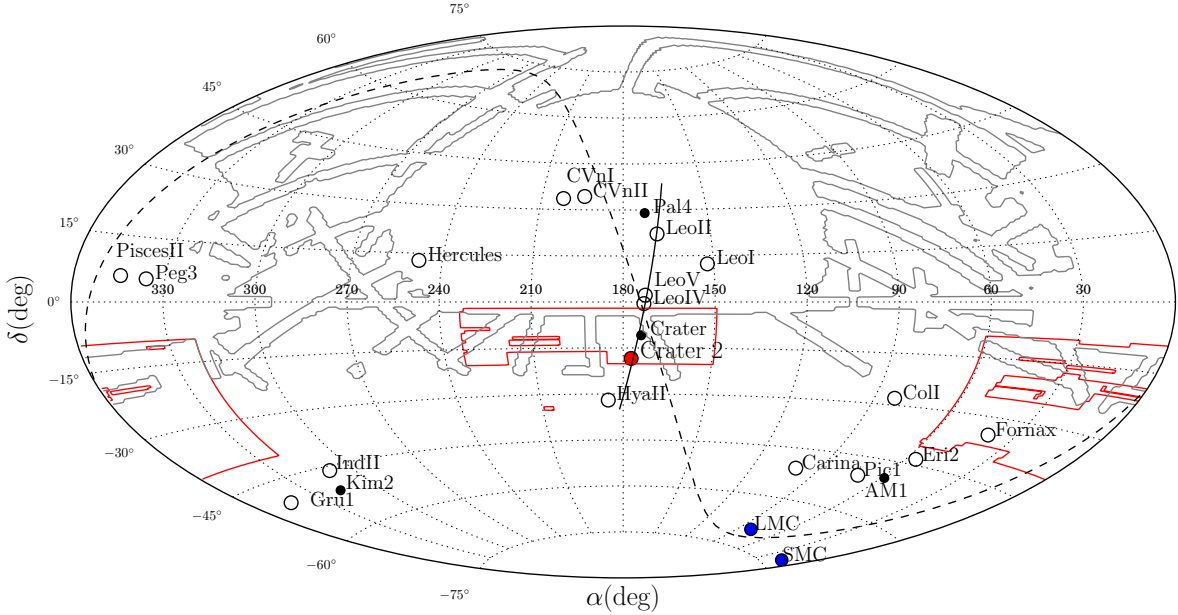}
    \caption{
      Positions of Milky Way satellites with heliocentric distances between 100\,kpc and 400\,kpc. Open circles are dwarf galaxies from \citet{McConnachie2012}, filled black circles are globular clusters, filled blue circles show the positions of the LMC and the SMC, and the filled red circle gives the position of Crater~2. The red (grey) contour delimits the ATLAS (SDSS) footprint. The black line shows the segment of the best fit orbit, which passes very close to the 3-D positions of Crater~2, Crater, Leo~IV, Leo~V and Leo~II. The great circle defined by \citet{Nidever2008} which aligns with the Magellanic stream is also shown with the black dashed line.}
    \label{fig:Crater2wholesky}
\end{figure*}

The location of the Crater~2 in the size-luminosity plane is rather remarkable. First, it is one of the lowest surface brightness dwarfs discovered so far, hovering just above the 31 mag\,arcsec$^{-2}$ mark. Comparably diffuse are the recently found Tuc~2 and IV as well as the pair of dwarfs in the constellation of Ursa Major. Second, the new dwarf appears to be the largest ultra-faint satellite known in the Galaxy, and the fourth largest overall: only the disrupting Sagittarius dwarf and the Magellanic Clouds - both with significantly larger stellar masses - are bigger than Crater~2.  In Andromeda on the other hand, it has long been suspected that there exists a sub-population of oversized dwarfs simply not present in the Milky Way. These would for example include And~XIX with the half-light radius of 1.7\,kpc \citep{Mcconnachie2008} and a somewhat brighter And XXXII with the half-light radius of 1.4\,kpc \citep{Martin2013}. At last, with the discovery of Crater~2 some semblance of parity has been established!

It is now being gradually acknowledged that a substantial number of very low surface brightness systems might have been avoiding detection. For example, recently, a discovery of an entire population of ultra diffuse galaxies (UDGs) was reported by \citet{vanDokkum2015} in the Coma cluster. Compared to Crater~2, the Coma UDGs are typically more extended, with sizes between 1.5 and ~4\,kpc. Most importantly, they reach much higher levels of surface brightness of $\sim$ 25 mag\,arcsec$^{-2}$, i.e. two orders of magnitude brighter than Crater~ 2 or Andromeda XIX. It is therefore difficult to claim an unambiguous evolutionary connection between the UDGs and dwarfs similar to Crater~2. For the UDGs, tides seem to be the likely culprit. However, as far as possible Crater~2 formation scenarios are concerned, we can confidently exclude interactions with the Galaxy. The two primary reasons being the near perfect circular shape of Crater~2 iso-density contours and its likely very long orbital period.

\begin{figure*}
    \includegraphics[width=\textwidth]{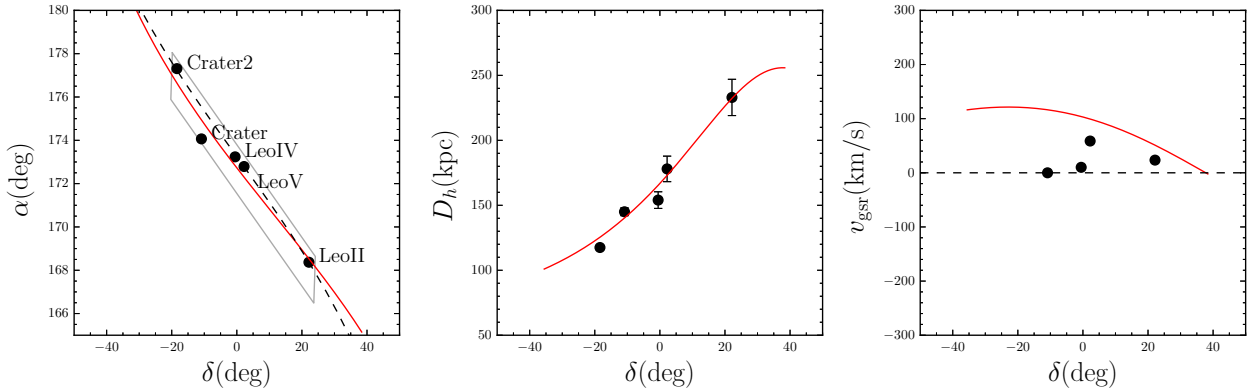}
    \caption{
      The Crater-Leo group. The left panel shows the positions of Crater, Crater~2, Leo~IV, Leo~V and Leo~II on the sky. Black dashed line indicates the great circle with the pole at $(\alpha,\delta)=(83.\degr2,-11.\degr8)$ that passes close to all five satellites, and the black box shows the size of the box used to estimate the significance of the association. The middle panel shows the run of the satellites' heliocentric distance versus declination, and the right panel shows the heliocentric radial velocity (corrected for the Solar reflex motion) versus declination. The red line shows an orbit obtained by modelling the 3-D positions and the velocities of the four satellites with known radial velocities, i.e., Leo~II, Leo~IV, Leo~V and Crater.}
    \label{fig:LeoCrater}
\end{figure*}

Perhaps the following simple, if mundane, genesis hypothesis is worth considering. Do systems like Crater~2 and And~XIX represent a natural continuation of the dwarf galaxy distribution in size and surface brightness? This could be the missing part of the dwarf family that only now comes into view helped by data from ever more ambitious wide-area surveys. Additionally, it is not very difficult to imagine that the bulk of the dwarf spheroidals were somewhat larger to begin with. Falling into the potential well of the Milky Way would lead to them experiencing varying amounts of tidal truncation \citep[see][]{Penarrubia2008}. If so then there should be a correlation between the size (and surface brightness) of a dwarf and the time since its in-fall, with the likes of Crater~2 representing the largely unaffected, recently accreted population. The question, however, remains whether the size and/or the surface brightness of a dwarf could be linked to the physical conditions at the birthplace. This argument can only be developed further if the orbital history of the satellite can be better constrained.

\subsection{The Crater-Leo group}

Figure \ref{fig:Crater2wholesky} shows the distribution on the sky (in equatorial coordinates) of all satellites (both dwarfs and globulars) located further than 100\,kpc from the Galactic center. \citet{Belokurov2010,Belokurov2014} have already pointed out the close alignment between Leo~IV, Leo~V and Crater. Not only are these positioned almost precisely along a great circle, they also possess very similar radial velocities and distances, suggestive of a common origin in a galaxy group prior to infall into the Milky Way. Surprisingly, as the Figure demonstrates, the newly discovered Crater~2 and the classical satellite Leo~II could also be part of this group: all five lie close to the great circle with the pole at $(\alpha,\delta)=(83.\degr2,-11.\degr8)$. In 3-D, as the middle panel of Figure~\ref{fig:LeoCrater} shows, the location of the satellites is consistent with a monotonic distance increase, moving up in declination from Crater~2 to Leo~II (bear in mind, though, that the heliocentric distance errors are considerable, i.e. around 10-15\%).

Let us investigate further the significance of the alignment between these distant satellites. We start by considering the Northern Galactic Cap (NGC) area (b$>$30\degr) as observed by both the VST ATLAS and the SDSS - together the two surveys have covered $\sim$9000 square degrees in this part of the sky. In this area, the total number of known satellites (globular clusters and dwarf galaxies) outside 100\,kpc radius is 10. Therefore, we create random Monte-Carlo realizations of satellite distribution on the sky within the northern SDSS and VST ATLAS footprints assuming uniform distribution on the sky, where the number of satellites is dictated by the Poisson distribution with $\lambda=10$. For each random realization, we determine the largest group of satellites that are strongly aligned, similarly to what is observed for the Crater-Leo group (the satellites in the Crater-Leo group fit within the great circle box with the size of $\sim$ 1.6 \degr $\times$ 42 \degr; see Fig.~\ref{fig:LeoCrater}). More specifically, for each simulation we find the maximum number of satellites that can fit within an elongated great circle box with the size 2\degr $\times$ 45\degr\ of any orientation and location on the sky. On average this number is $\sim$ 2.8, however having aligned groups with 5 or more members is very unlikely, the probability of this is $\sim$ 0.0045 which corresponds to $\sim$ 2.6$\sigma$. Even though the significance appears rather modest, note that this calculation already takes into account the ``spatial look-elsewhere'' effect, as in the simulations we look for possible alignments at any position on the sky and in any orientation. It is also important that the calculation of the significance has been done assuming a null-hypothesis of uniform distribution of satellites on the sky. If it is not uniform and and is dominated by a plane of satellites as argued by e.g. \citep{Pawlowski2012}, the significance of the Crater-Leo group will be different and will depend on detailed properties of the plane of satellites.

The true significance of the orbital synchronization is probably still higher given that both their distances (radial velocities) are similar for the 5 (4) objects in the Leo/Crater group. To assess the plausibility of the group accretion hypothesis we attempt to model the positions and the kinematics of the Crater-Leo satellites with a single orbit. We used a Navarro-Frenk-White (NFW) halo with a virial mass of 10$^{12}$\,M$_\odot$ and concentration $c=10$. The best-fitting orbit that reproduces the 3-D positions and goes close to the radial velocities of the satellites is given in Figure~\ref{fig:LeoCrater}. It is, however, immediately clear from the inspection of the Figure, that although a single orbit can describe the 3-D positions well, it can not simultaneously match the observed kinematics.

The poor kinematics fit is not surprising: low amplitude of radial motion (as manifested in all four satellites with known velocities) corresponds to the orbital extrema, the pericentre and the apocentre. The kinematics would therefore require these four group members to cluster in the vicinity of either apo or peri, while in reality their distances cover an impressive range of at least 100\,kpc. It is, however, well known that tidal debris do not follow orbits exactly. There exists strong energy sorting of debris along the stream, imprinted at the time of stripping. The energy offset between the debris and the progenitor is a function of orbital phase \citep[see e.g.][]{Gibbons2014} - the widest energy distribution available around the peri crossing, with the objects unbound closer to the apo-centre characterized by energies more similar to that of the progenitor. Thus, it is not unreasonable to conjecture that the satellites stripped from the common group and that are now observed at different phases along the stream, follow distinct orbits, corresponding to the energy at the time of un-binding, each with its own apo-centre.

As a matter of fact, long and approximately linear tidal ``streamlets'' reaching well beyond the average stream apo-centre are a ubiquitous feature in simulations of the Sgr dwarf disruption. These, as discussed above, are composed of high energy particles torn off the progenitor at the penultimate peri-centre crossing. If Crater~2 indeed came from the same progenitor, to which we assume Crater, Leo IV, V and II all one day belonged, and the group is currently near the apo-centre, the amplitude of the line-of-sight velocity of Crater~2 should be low. Ultimately, to confirm this hypothesis one would require accurate proper motion measurements for as many members of the tentative group as possible. So far, such measurement only exists for the Leo~II dwarf \citep{Lepine2011}. Unfortunately, the measurement error is large and only loosely constrains the direction of motion of Leo~II, which appears to be marginally consistent with the plane that contains the best-fit orbit.

Note that there are further, perhaps rather circumstantial, pieces of evidence in favor of the Crater-Crater~2 connection. Both objects are amongst the most extended in the corresponding satellite category: Crater~2 amongst the Milky Way dwarfs, and Crater amongst the Milky Way globular clusters. Both are highly spherical. Finally, in terms of stellar populations, both have red and short horizontal branches, whose appearance is somewhat unusual given the objects' low metallicity.

\subsection{Summary}

Here we have presented the discovery of the Crater~2 dwarf galaxy, identified in the data of the VST ATLAS survey. The luminosity of Crater~2 is $M_V\approx-8$, i.e. in the range where it could be classified both as a bright ultra-faint dwarf or as a faint classical one. With the half-light radius of $\sim$1.1\,kpc, Crater~2 would be the largest amongst the UFDs, and is the fourth largest amongst the Milky Way satellites overall - only the LMC, the SMC and the Sgr are more extended. Consequently, Crater~2 is also one of the lowest surface brightness stellar systems in the Universe ever discovered. Additionally, there is a mounting body of evidence that Crater~2 could have belonged to the satellite group, whose members also included the peculiar globular cluster Crater, UFDs Leo~IV and V, as well as the classical dwarf Leo~II. We hypothesize that the Crater-Leo group is now dissolved into a narrow tidal debris stream whose apo-centre is between 100 and 200\,kpc. The stellar overdensity corresponding to Crater~2 was detected at comfortably high levels of statistical significance. If similar size objects with lower luminosity - and hence lower surface brightness - exist, they should be uncovered in the nearest future.

\section*{Acknowledgments} 

The authors wish to acknowledge many useful discussions with the members of the Cambridge STREAMS club in general, and Wyn Evans, Denis Erkal and Simon Gibbons in particular. We also are grateful to Filippo Fraternali for his advice on the HI data available, and to the anonymous referee for detailed and useful comments. Support for G.T. is provided by CONICYT Chile. The research leading to these results has received funding from the European Research Council under the European Union's Seventh Framework Programme (FP/2007-2013)/ERC Grant Agreement no. 308024. This research was made possible through the use of the AAVSO Photometric All-Sky Survey (APASS), funded by the Robert Martin Ayers Sciences Fund.

\bibliographystyle{mn2e}
\bibliography{biblio}

\begin{thebibliography}{51}
\expandafter\ifx\csname natexlab\endcsname\relax\def\natexlab#1{#1}\fi

\bibitem[{{Balbinot} {et~al}\mbox{.}(2013){Balbinot}, {Santiago}, {da Costa},
  {Maia}, {Majewski}, {Nidever}, {Rocha-Pinto}, {Thomas}, {Wechsler}, \&
  {Yanny}}]{Balbinot2013}
{Balbinot} E. {et~al.}, 2013, \apj, 767, 101

\bibitem[{{Bechtol} {et~al}\mbox{.}(2015){Bechtol}, {Drlica-Wagner},
  {Balbinot}, {Pieres}, {Simon}, {Yanny}, {Santiago}, {Wechsler}, {Frieman},
  {Walker}, {Williams}, {Rozo}, {Rykoff}, \& {The DES
  Collaboration}}]{Bechtol2015}
{Bechtol} K. {et~al.}, 2015, \apj, 807, 50

\bibitem[{{Belokurov} {et~al}\mbox{.}(2014){Belokurov}, {Irwin}, {Koposov},
  {Evans}, {Gonzalez-Solares}, {Metcalfe}, \& {Shanks}}]{Belokurov2014}
{Belokurov} V., {Irwin} M.~J., {Koposov} S.~E., {Evans} N.~W.,
  {Gonzalez-Solares} E., {Metcalfe} N., {Shanks} T., 2014, \mnras, 441, 2124

\bibitem[{{Belokurov} {et~al}\mbox{.}(2008){Belokurov}, {Walker}, {Evans},
  {Faria}, {Gilmore}, {Irwin}, {Koposov}, {Mateo}, {Olszewski}, \&
  {Zucker}}]{Belokurov2008}
{Belokurov} V. {et~al.}, 2008, \apjl, 686, L83

\bibitem[{{Belokurov} {et~al}\mbox{.}(2010){Belokurov}, {Walker}, {Evans},
  {Gilmore}, {Irwin}, {Just}, {Koposov}, {Mateo}, {Olszewski}, {Watkins}, \&
  {Wyrzykowski}}]{Belokurov2010}
{Belokurov} V. {et~al.}, 2010, \apjl, 712, L103

\bibitem[{{Belokurov} {et~al}\mbox{.}(2007){Belokurov}, {Zucker}, {Evans},
  {et~al.}}]{Belokurov2007}
{Belokurov} V., {Zucker} D.~B., {Evans} N.~W., {et~al.}, 2007, \apj, 654, 897

\bibitem[{{Bressan} {et~al}\mbox{.}(2012){Bressan}, {Marigo}, {Girardi},
  {Salasnich}, {Dal Cero}, {Rubele}, \& {Nanni}}]{Bressan2012}
{Bressan} A., {Marigo} P., {Girardi} L., {Salasnich} B., {Dal Cero} C.,
  {Rubele} S., {Nanni} A., 2012, \mnras, 427, 127

\bibitem[{{Brodie} {et~al}\mbox{.}(2011){Brodie}, {Romanowsky}, {Strader}, \&
  {Forbes}}]{Brodie2011}
{Brodie} J.~P., {Romanowsky} A.~J., {Strader} J., {Forbes} D.~A., 2011, \aj,
  142, 199

\bibitem[{{Chabrier}(2003)}]{Chabrier2003}
{Chabrier} G., 2003, \pasp, 115, 763

\bibitem[{{Chen}, {Zhao} \& {Zhao}(2009){Chen}, {Zhao}, \& {Zhao}}]{Chen2009}
{Chen} Y.~Q., {Zhao} G., {Zhao} J.~K., 2009, \apj, 702, 1336

\bibitem[{{Deason} {et~al}\mbox{.}(2015){Deason}, {Wetzel}, {Garrison-Kimmel},
  \& {Belokurov}}]{Deason2015}
{Deason} A.~J., {Wetzel} A.~R., {Garrison-Kimmel} S., {Belokurov} V., 2015,
  \mnras, 453, 3568

\bibitem[{{Drlica-Wagner} {et~al}\mbox{.}(2015){Drlica-Wagner}, {Bechtol},
  {Rykoff}, {Luque}, {Queiroz}, {Mao}, {Wechsler}, {Simon}, \& {The DES
  Collaboration}}]{DrlicaWagner2015}
{Drlica-Wagner} A. {et~al.}, 2015, \apj, 813, 109

\bibitem[{{Foreman-Mackey} {et~al}\mbox{.}(2013){Foreman-Mackey}, {Hogg},
  {Lang}, \& {Goodman}}]{ForemanMackey2013}
{Foreman-Mackey} D., {Hogg} D.~W., {Lang} D., {Goodman} J., 2013, \pasp, 125,
  306

\bibitem[{{Gibbons}, {Belokurov} \& {Evans}(2014){Gibbons}, {Belokurov}, \&
  {Evans}}]{Gibbons2014}
{Gibbons} S.~L.~J., {Belokurov} V., {Evans} N.~W., 2014, \mnras, 445, 3788

\bibitem[{{Goodman} \& {Weare}(2010)}]{GoodmanWeare2010}
{Goodman} J., {Weare} J., 2010, Comm. App. Math. Comp. Sci, 5, 65

\bibitem[{{Gratton} {et~al}\mbox{.}(2010){Gratton}, {Carretta}, {Bragaglia},
  {Lucatello}, \& {D'Orazi}}]{Gratton2010}
{Gratton} R.~G., {Carretta} E., {Bragaglia} A., {Lucatello} S., {D'Orazi} V.,
  2010, \aap, 517, A81

\bibitem[{{Harris}(2010)}]{Harris2010}
{Harris} W.~E., 2010, ArXiv e-prints

\bibitem[{{Ibata}, {Gilmore} \& {Irwin}(1994){Ibata}, {Gilmore}, \&
  {Irwin}}]{Ibata1994}
{Ibata} R.~A., {Gilmore} G., {Irwin} M.~J., 1994, \nat, 370, 194

\bibitem[{{Irwin} {et~al}\mbox{.}(1990){Irwin}, {Bunclark}, {Bridgeland}, \&
  {McMahon}}]{Sextans}
{Irwin} M.~J., {Bunclark} P.~S., {Bridgeland} M.~T., {McMahon} R.~G., 1990,
  \mnras, 244, 16P

\bibitem[{{Kim} \& {Jerjen}(2015)}]{Kim2015a}
{Kim} D., {Jerjen} H., 2015, \apj, 799, 73

\bibitem[{{Kim} {et~al}\mbox{.}(2015{\natexlab{a}}){Kim}, {Jerjen}, {Mackey},
  {Da Costa}, \& {Milone}}]{Kim2015c}
{Kim} D., {Jerjen} H., {Mackey} D., {Da Costa} G.~S., {Milone} A.~P.,
  2015{\natexlab{a}}, ArXiv e-prints

\bibitem[{{Kim} {et~al}\mbox{.}(2015{\natexlab{b}}){Kim}, {Jerjen}, {Milone},
  {Mackey}, \& {Da Costa}}]{Kim2015b}
{Kim} D., {Jerjen} H., {Milone} A.~P., {Mackey} D., {Da Costa} G.~S.,
  2015{\natexlab{b}}, \apj, 803, 63

\bibitem[{{Koposov} {et~al}\mbox{.}(2008){Koposov}, {Belokurov}, {Evans},
  {Hewett}, {Irwin}, {Gilmore}, {Zucker}, {Rix}, {Fellhauer}, {Bell}, \&
  {Glushkova}}]{Koposov2008}
{Koposov} S. {et~al.}, 2008, \apj, 686, 279

\bibitem[{{Koposov} {et~al}\mbox{.}(2015){Koposov}, {Belokurov}, {Torrealba},
  \& {Evans}}]{Koposov2015}
{Koposov} S.~E., {Belokurov} V., {Torrealba} G., {Evans} N.~W., 2015, \apj,
  805, 130

\bibitem[{{Koposov} {et~al}\mbox{.}(2014){Koposov}, {Irwin}, {Belokurov},
  {Gonzalez-Solares}, {Yoldas}, {Lewis}, {Metcalfe}, \& {Shanks}}]{Koposov2014}
{Koposov} S.~E., {Irwin} M., {Belokurov} V., {Gonzalez-Solares} E., {Yoldas}
  A.~K., {Lewis} J., {Metcalfe} N., {Shanks} T., 2014, \mnras, 442, L85

\bibitem[{{Kroupa}, {Theis} \& {Boily}(2005){Kroupa}, {Theis}, \&
  {Boily}}]{Kroupa2005}
{Kroupa} P., {Theis} C., {Boily} C.~M., 2005, \aap, 431, 517

\bibitem[{{Laevens} {et~al}\mbox{.}(2015){Laevens}, {Martin}, {Bernard},
  {Schlafly}, {Sesar}, {Rix}, {Bell}, {Ferguson}, {Slater}, {Sweeney}, {Wyse},
  {Huxor}, {Burgett}, {Chambers}, {Draper}, {Hodapp}, {Kaiser}, {Magnier},
  {Metcalfe}, {Tonry}, {Wainscoat}, \& {Waters}}]{Laevens2015}
{Laevens} B.~P.~M. {et~al.}, 2015, \apj, 813, 44

\bibitem[{{Laevens} {et~al}\mbox{.}(2014){Laevens}, {Martin}, {Sesar},
  {Bernard}, {Rix}, {Slater}, {Bell}, {Ferguson}, {Schlafly}, {Burgett},
  {Chambers}, {Denneau}, {Draper}, {Kaiser}, {Kudritzki}, {Magnier},
  {Metcalfe}, {Morgan}, \& {Price}}]{Laevens2014}
{Laevens} B.~P.~M. {et~al.}, 2014, \apjl, 786, L3

\bibitem[{{L{\'e}pine} {et~al}\mbox{.}(2011){L{\'e}pine}, {Koch}, {Rich}, \&
  {Kuijken}}]{Lepine2011}
{L{\'e}pine} S., {Koch} A., {Rich} R.~M., {Kuijken} K., 2011, \apj, 741, 100

\bibitem[{{Lynden-Bell}(1976)}]{Lyndenbell1976}
{Lynden-Bell} D., 1976, \mnras, 174, 695

\bibitem[{{Lynden-Bell} \& {Lynden-Bell}(1995)}]{LyndenBell1995}
{Lynden-Bell} D., {Lynden-Bell} R.~M., 1995, \mnras, 275, 429

\bibitem[{{Martin}, {de Jong} \& {Rix}(2008){Martin}, {de Jong}, \&
  {Rix}}]{Martin2008}
{Martin} N.~F., {de Jong} J.~T.~A., {Rix} H.-W., 2008, \apj, 684, 1075

\bibitem[{{Martin} {et~al}\mbox{.}(2013){Martin}, {Slater}, {Schlafly},
  {Morganson}, {Rix}, {Bell}, {Laevens}, {Bernard}, {Ferguson}, {Finkbeiner},
  {Burgett}, {Chambers}, {Hodapp}, {Kaiser}, {Kudritzki}, {Magnier}, {Morgan},
  {Price}, {Tonry}, \& {Wainscoat}}]{Martin2013}
{Martin} N.~F. {et~al.}, 2013, \apj, 772, 15

\bibitem[{{McConnachie}(2012)}]{McConnachie2012}
{McConnachie} A.~W., 2012, \aj, 144, 4

\bibitem[{{McConnachie} {et~al}\mbox{.}(2008){McConnachie}, {Huxor}, {Martin},
  {Irwin}, {Chapman}, {Fahlman}, {Ferguson}, {Ibata}, {Lewis}, {Richer}, \&
  {Tanvir}}]{Mcconnachie2008}
{McConnachie} A.~W. {et~al.}, 2008, \apj, 688, 1009

\bibitem[{{Mu{\~n}oz} {et~al}\mbox{.}(2012){Mu{\~n}oz}, {Geha}, {C{\^o}t{\'e}},
  {Vargas}, {Santana}, {Stetson}, {Simon}, \& {Djorgovski}}]{Munoz2012}
{Mu{\~n}oz} R.~R., {Geha} M., {C{\^o}t{\'e}} P., {Vargas} L.~C., {Santana}
  F.~A., {Stetson} P., {Simon} J.~D., {Djorgovski} S.~G., 2012, \apjl, 753, L15

\bibitem[{{Nichols} {et~al}\mbox{.}(2011){Nichols}, {Colless}, {Colless}, \&
  {Bland-Hawthorn}}]{Nichols2011}
{Nichols} M., {Colless} J., {Colless} M., {Bland-Hawthorn} J., 2011, \apj, 742,
  110

\bibitem[{{Nidever}, {Majewski} \& {Burton}(2008){Nidever}, {Majewski}, \&
  {Burton}}]{Nidever2008}
{Nidever} D.~L., {Majewski} S.~R., {Burton} W.~B., 2008, \apj, 679, 432

\bibitem[{{Niederste-Ostholt} {et~al}\mbox{.}(2010){Niederste-Ostholt},
  {Belokurov}, {Evans}, \& {Pe{\~n}arrubia}}]{MNO2010}
{Niederste-Ostholt} M., {Belokurov} V., {Evans} N.~W., {Pe{\~n}arrubia} J.,
  2010, \apj, 712, 516

\bibitem[{{Padmanabhan} {et~al}\mbox{.}(2008){Padmanabhan}, {Finkbeiner},
  {Barentine}, {Blanton}, {Brewington}, {Gunn}, \& {Harvanek}}]{Padma08}
{Padmanabhan}, N.and~{Schlegel} D.~J., {Finkbeiner} D.~P., {Barentine} J.~C.,
  {Blanton} M.~R., {Brewington} H.~J., {Gunn} J.~E., {Harvanek} M., 2008, \apj,
  674, 1217

\bibitem[{{Pawlowski}, {Pflamm-Altenburg} \& {Kroupa}(2012){Pawlowski},
  {Pflamm-Altenburg}, \& {Kroupa}}]{Pawlowski2012}
{Pawlowski} M.~S., {Pflamm-Altenburg} J., {Kroupa} P., 2012, \mnras, 423, 1109

\bibitem[{{Pe{\~n}arrubia}, {Navarro} \& {McConnachie}(2008){Pe{\~n}arrubia},
  {Navarro}, \& {McConnachie}}]{Penarrubia2008}
{Pe{\~n}arrubia} J., {Navarro} J.~F., {McConnachie} A.~W., 2008, \apj, 673, 226

\bibitem[{{Sales} {et~al}\mbox{.}(2011){Sales}, {Navarro}, {Cooper}, {White},
  {Frenk}, \& {Helmi}}]{Sales2011}
{Sales} L.~V., {Navarro} J.~F., {Cooper} A.~P., {White} S.~D.~M., {Frenk}
  C.~S., {Helmi} A., 2011, \mnras, 418, 648

\bibitem[{{Schlegel}, {Finkbeiner} \& {Davis}(1998){Schlegel}, {Finkbeiner}, \&
  {Davis}}]{SFD}
{Schlegel} D.~J., {Finkbeiner} D.~P., {Davis} M., 1998, \apj, 500, 525

\bibitem[{{Shanks} {et~al}\mbox{.}(2015){Shanks}, {Metcalfe}, {Chehade},
  {Findlay}, {Irwin}, {Gonzalez-Solares}, {Lewis}, {Yoldas}, {Mann}, {Read},
  {Sutorius}, \& {Voutsinas}}]{Shanks2015}
{Shanks} T. {et~al.}, 2015, \mnras, 451, 4238

\bibitem[{{Skrutskie} {et~al}\mbox{.}(2006){Skrutskie}, {Cutri}, {Stiening},
  {Weinberg}, {Schneider}, {Carpenter}, {Beichman}, {Capps}, {Chester},
  {Elias}, {Huchra}, {Liebert}, {Lonsdale}, {Monet}, {Price}, {Seitzer},
  {Jarrett}, {Kirkpatrick}, {Gizis}, {Howard}, {Evans}, {Fowler}, {Fullmer},
  {Hurt}, {Light}, {Kopan}, {Marsh}, {McCallon}, {Tam}, {Van Dyk}, \&
  {Wheelock}}]{Skrutskie2006}
{Skrutskie} M.~F. {et~al.}, 2006, \aj, 131, 1163

\bibitem[{{van Dokkum} {et~al}\mbox{.}(2015){van Dokkum}, {Abraham}, {Merritt},
  {Zhang}, {Geha}, \& {Conroy}}]{vanDokkum2015}
{van Dokkum} P.~G., {Abraham} R., {Merritt} A., {Zhang} J., {Geha} M., {Conroy}
  C., 2015, \apjl, 798, L45

\bibitem[{{Weisz} {et~al}\mbox{.}(2015){Weisz}, {Koposov}, {Dolphin},
  {Belokurov}, {Gieles}, {Mateo}, {Olszewski}, {Sills}, \&
  {Walker}}]{Weisz2015}
{Weisz} D.~R. {et~al.}, 2015, ArXiv e-prints

\bibitem[{{Wetzel}, {Deason} \& {Garrison-Kimmel}(2015){Wetzel}, {Deason}, \&
  {Garrison-Kimmel}}]{Wetzel2015}
{Wetzel} A.~R., {Deason} A.~J., {Garrison-Kimmel} S., 2015, \apj, 807, 49

\bibitem[{{Yozin} \& {Bekki}(2015)}]{Yozin2015}
{Yozin} C., {Bekki} K., 2015, \mnras, 453, 2302

\bibitem[{{Yuan}, {Liu} \& {Xiang}(2013){Yuan}, {Liu}, \& {Xiang}}]{Yuan2013}
{Yuan} H.~B., {Liu} X.~W., {Xiang} M.~S., 2013, \mnras, 430, 2188

\end{thebibliography}

\label{lastpage}

\end{document}